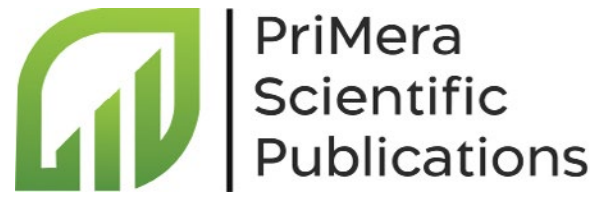

# Semantic Reverse Engineering Legacy Software Applications with ChatGPT, Gemini AI, and Claude AI

**Christian Mancas* and Diana Christina Mancas**

*Mathematics and Computer Science Department, Ovidius University at Constanta, Romania*

***Corresponding Author:** Christian Mancas, Ovidius University, Bd. Mamaia 124, Constanta, CT, Romania.

## Abstract

This research paper describes our research results on using ChatGPT, Gemini, and Claude AI to semantically reverse engineer legacy database software applications.

## Introduction

In our previously published paper [1], we reported our research on using ChatGPT [2], Gemini AI [3], Claude AI [4], and *MatBase* [5] for reverse engineering legacy databases (dbs) into our (Elementary) Mathematical Data Model ((E)MDM) [6] schemas. Database (db) reverse engineering (re) is only the first step in the re of db software applications (apps). This paper reports on our latest research focused on apps re, using the power of these Artificial Intelligence (AI) tools.

As an example, we chose code snippets from the *Geography* db app provided by *MatBase* and from a *Genealogy* one that are both enforcing non-relational constraints. *MatBase* is our intelligent data and knowledge base management system prototype [5, 7] based on both (E)MDM (which includes Datalog¬[8]), the Relational (RDM) [8-10] and Entity-Relationship (E-R) [10-12] Data Models. *MatBase's* main goal is to provide modeling as programming [13] and, especially, mathematical data modeling. *MatBase* has two versions, one for small and medium dbs developed in MS VBA over Access, and one for large dbs developed in MS C# over SQL Server.

*MatBase* automatically generates VBA, C#, and SQL code for enforcing most of the 78 (E)MDM constraint types but does not have software re capabilities. This is why we used for this research only the AI tools ChatGPT, Gemini, and Claude.

Generally, software re is purely syntactic, i.e., striving only to obtain the source code, not its intent. Our aim is a much higher, semantic one: given source legacy code (generally, enforcing non-relational db constraints), reverse engineer it to infer a formal definition of the corresponding constraints, using the naïve theory of sets, relations, and functions [14], as well as the first order predicate logic with equality [15], plus its temporal extension [16], and then also accurately express it informally in plain English.

The next Sections explore related work, present the materials and methods used, the results obtained and discussed, conclusions, and the reference list.

***Related Work***

Although we too devoted throughout our career most of the time to software forward engineering, we also considered re, its dual, not only in [17], but also in the 4th Chapter of [10]. The seminal paper in this field remains [18].

Software re (sre) is a vast field, incorporating binary code re, including the volatile RAM stored one [19], not only the file stored one [20], malware detection and protection [19, 21], the famous NSA Ghidra Suite [22], etc.. Most of them are dedicated to an operating system family, from Windows [23] to mobile and IoT ones [24]. For the MS Visual Basic for Application (VBA), besides malware detection and protection [25], the issue addressed is generally code extracting and deobfuscation, for its both pseudo and native code [26]. From all these tools, only VB Decompiler [26] is advertised as also using AI, to increase its otherwise maximum recovery rates of 75% for native and 85% for pseudo code (see https://www.vb-decompiler.org/). All such theories and tools are purely syntactic.

## Materials and Methods

For this research, we used two Toshiba Satellite Intel CORE i7, MS Windows 10, Google Chrome version 146.0.7680.81, MS Access 365 v. 2603, *MatBase* v. 5.2 Access, ChatGPT Plus 5.3, Gemini 3, and Claude Sonnet 4.6.

The architecture of the MS Access *MatBase* version [7] is a standard one: the MatBase.mdb pure VBA code one (i.e., containing only forms, reports, VBA code for enforcing non-relational constraints, SQL queries, and a menu) uses links to the tables of several pure data dbs - some storing fundamental data, like MatBaseDB.accdb for its metacatalog, GeographyDB.mdb, BookstoreDB.mdb, StocksDB.mdb, and UserDB.mdb for its corresponding example apps, as well as several dual ones storing only temporary tables, like MatBaseTmp.accdb, GeographyTmp.mdb, BookstoreTmp.mdb, StocksTmp.mdb, and UserTmp.mdb.

The MatBase.mdb file contains 2144 objects (1449 queries, 342 forms, 2 reports, 15 modules, 1 local table (storing its table-driven menu form), 112 linked tables, and 223 other objects, like macros, dependencies, relationships, etc.) storing almost 26MB of data. The MS Access Database Documenter could not generate for this app a .pdf file of all its objects: Figure 1 shows the corresponding error message. However, we managed to obtain 6255 A4 pages of .pdf documentation (taking 24MB+) for the 342 forms and other 329 pages (taking 1.4MB+) for the 15 modules.

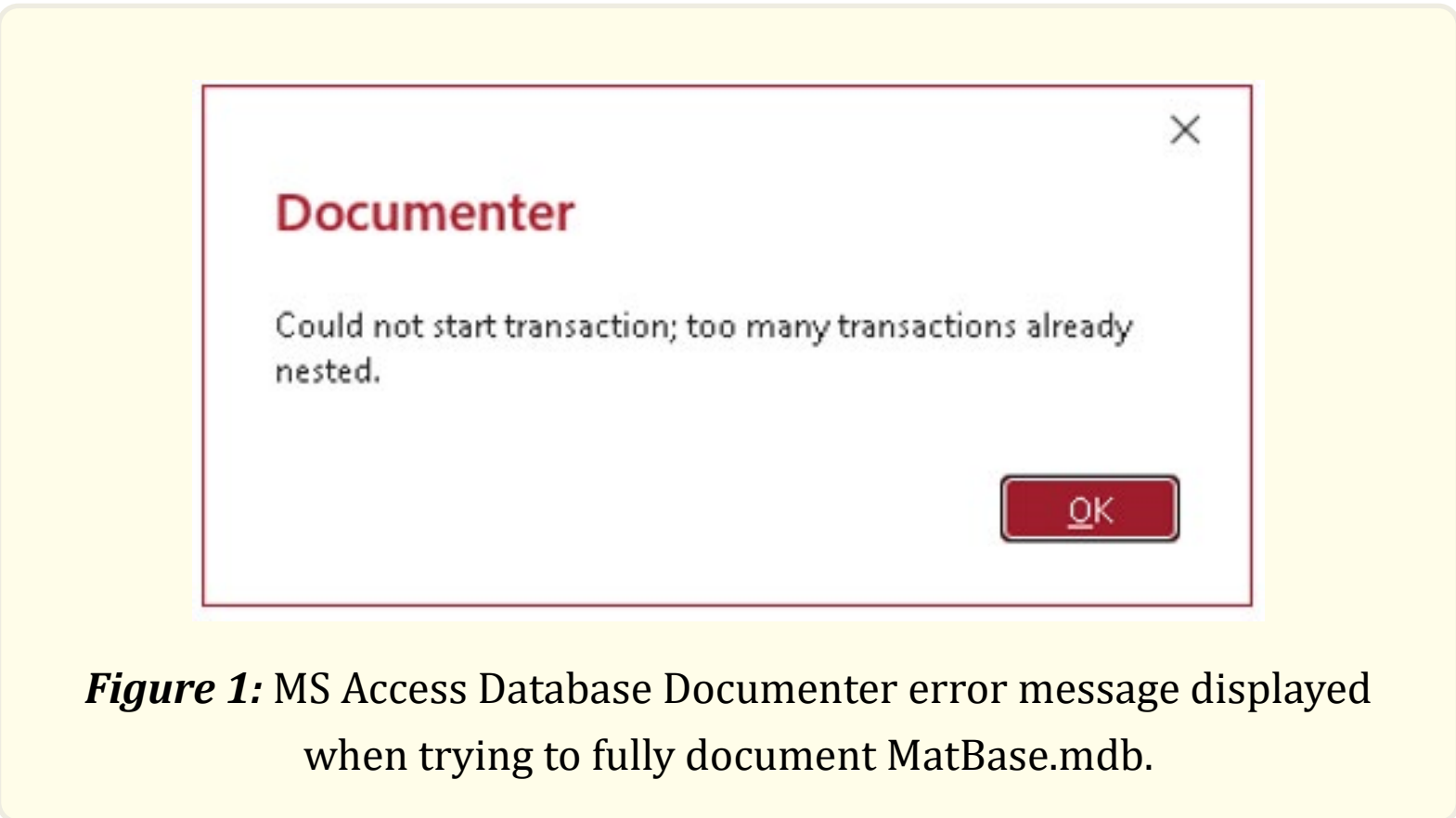

***Figure 1:*** MS Access Database Documenter error message displayed when trying to fully document MatBase.mdb.

MS Windows forms are bidimensional: they are made of a Graphic User Interface (GUI) set of controls, each having their properties, and of a class of event-driven and/or object-oriented VBA methods associated with them. Typically, the documentation generated for such a form has some 12 A4 pages of controls' properties and none to dozens of pages for the associated VBA class. Figure 2 shows a fragment of GUI documentation. The MS Access modules are VBA libraries, for storing constants, variables, and methods that are public to the apps. Figure 3 shows a fragment of a VBA code documentation (which is identically formatted for both form classes and modules).

Obviously, besides the SQL code for *RowSource* type properties of GUI combo box-type controls (see, e.g., the top of Figure 2), GUI documentation is not at all interesting for sre. Dually, the VBA embedding SQL code constitutes 99.95% of the semantic sre input. Although *MatBase* code includes lot of comments (see, e.g., in Figure 3 all texts prefixed by an apostrophe), we chose to eliminate them from our inputs given to the AI tools, as, generally, this is the case for the vast majority of legacy code, and being also curious to test their maximum level of code understanding and semantic abstraction power. We first illustrate our approach with two VBA code snippets enforcing the non-relational constraint corresponding to the "business" rule "Any country capital must be a city belonging to that country".

For example, Figure 4 shows the start of our dialogue with ChatGPT for this research. Note, please, that, on one hand, we eliminated all comments and did not give any info on the underlying db structure but, on the other, we did not eliminate the statement displaying error messages.

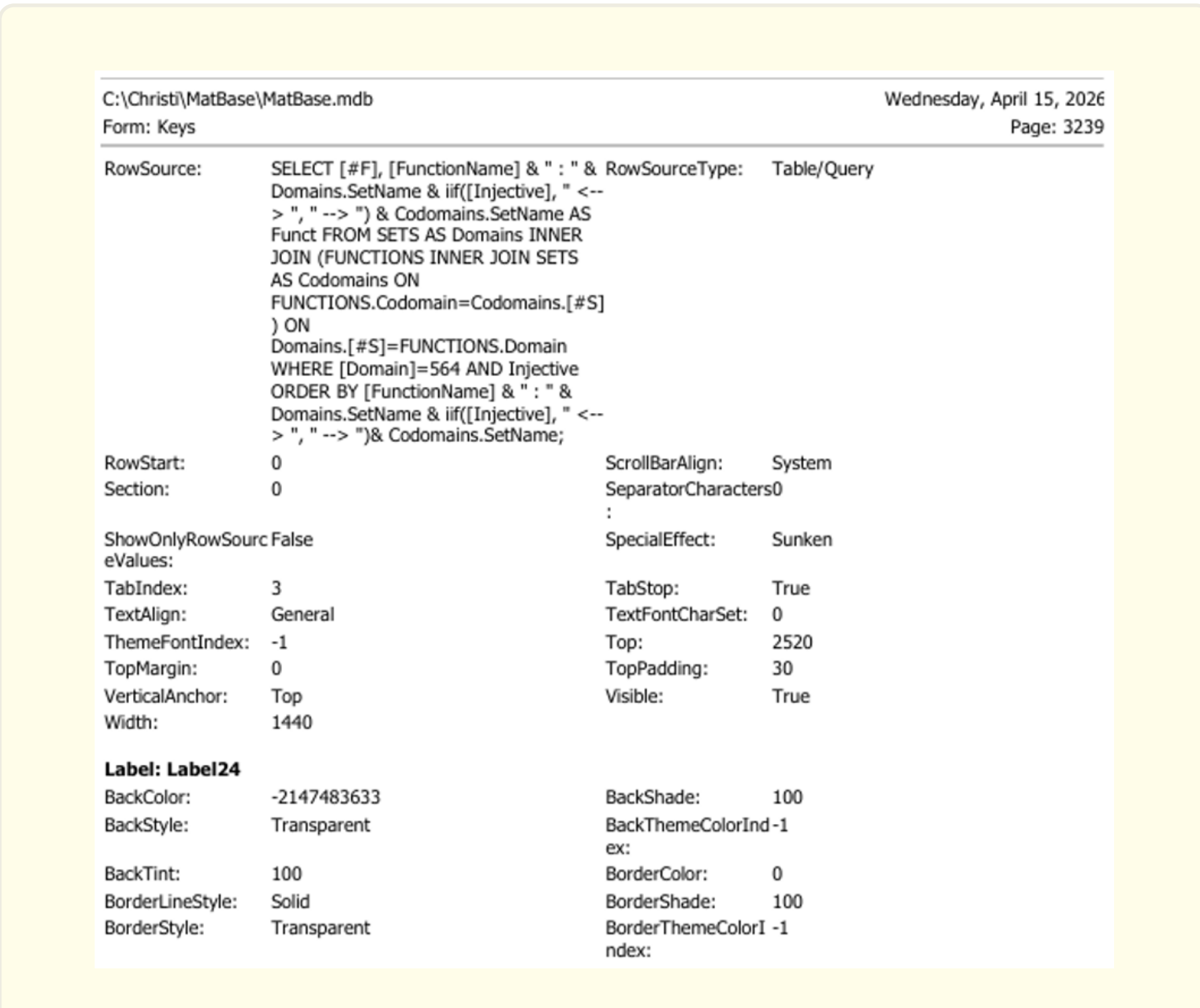

```
C:\Christi\MatBase\MatBase.mdb                         Wednesday, April 15, 2026
Form: Keys                                                            Page: 3239

RowSource:        SELECT [#F], [FunctionName] & " : " &   RowSourceType:     Table/Query
                  Domains.SetName & iif([Injective], " <--
                  > ", " --> ") & Codomains.SetName AS
                  Funct FROM SETS AS Domains INNER
                  JOIN (FUNCTIONS INNER JOIN SETS
                  AS Codomains ON
                  FUNCTIONS.Codomain=Codomains.[#S]
                  ) ON
                  Domains.[#S]=FUNCTIONS.Domain
                  WHERE [Domain]=564 AND Injective
                  ORDER BY [FunctionName] & " : " &
                  Domains.SetName & iif([Injective], " <--
                  > ", " --> ")& Codomains.SetName;
RowStart:         0                                       ScrollBarAlign:    System
Section:          0                                       SeparatorCharacters0
                                                          :
ShowOnlyRowSourc  False                                   SpecialEffect:     Sunken
eValues:
TabIndex:         3                                       TabStop:           True
TextAlign:        General                                 TextFontCharSet:   0
ThemeFontIndex:   -1                                      Top:               2520
TopMargin:        0                                       TopPadding:        30
VerticalAnchor:   Top                                     Visible:           True
Width:            1440

Label: Label24
BackColor:        -2147483633                             BackShade:         100
BackStyle:        Transparent                             BackThemeColorInd  -1
                                                          ex:
BackTint:         100                                     BorderColor:       0
BorderLineStyle:  Solid                                   BorderShade:       100
BorderStyle:      Transparent                             BorderThemeColorI  -1
                                                          ndex:
```

***Figure 2:*** MS Access Database Documenter fragment of a form GUI documentation.

## Results and Discussion

### *ChatGPT*

From Figure 5, which shows the beginning of its answer, it is clear that, immediately, ChatGPT understood the code and correctly stated the "business" rule it is enforcing: "A capital city must remain in its assigned country". Then, ChatGPT provided explanations line-by-line, giving proofs of its deep code intent understanding: e.g., for the line “If Not NewRecord Then”, it wrote “This means: Only check existing records. Not: New city entries. Because new cities aren’t capitals yet”. Similarly, for the line “cntry = Dlookup(…”, it correctly wrote “Find which country has this city as its capital”, although it did not guess that x is the name of the form control corresponding to the surrogate autonumber primary key of table *CITIES* but thought it is an undeclared alphanumeric variable. Consequently, it wrongly rewrote the code, assuming that *Capital* is alphanumeric too, without “thinking” that, e.g., besides the capital of France, there are 7 “Paris” U.S. cities (plus several other homonym towns) - hence, city names may not uniquely identify them (errors made then by Gemini and Claude as well).

Figure 6 shows the formalization of this “business” rule that follows, which is correct. Then, ChatGPT even correctly guesses the underlying db schema and the fact that this constraint may not be enforced by foreign keys, so it is a non-relational one (it calls it both an “application-level”, “runtime integrity”, and “semantic” constraint). The answer end, which is also correct, except for its middle part, where ChatGPT hallucinates on the corresponding function diagram (it is a circular, not a commutative type one [27]) is shown in Figure 7.

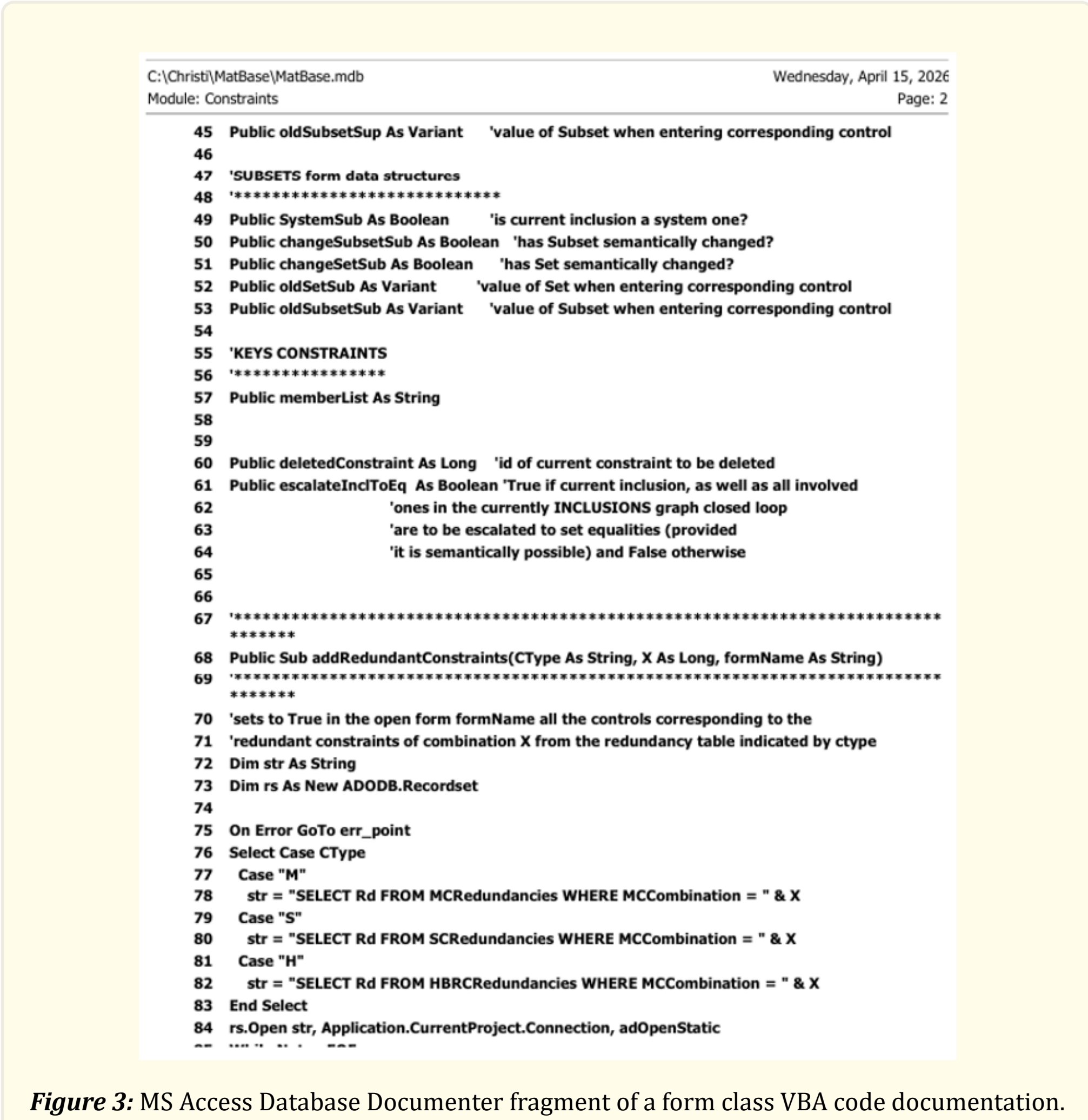

```
C:\Christi\MatBase\MatBase.mdb                                   Wednesday, April 15, 2026
Module: Constraints                                                               Page: 2
45  Public oldSubsetSup As Variant     'value of Subset when entering corresponding control
46
47  'SUBSETS form data structures
48  '***************************
49  Public SystemSub As Boolean        'is current inclusion a system one?
50  Public changeSubsetSub As Boolean   'has Subset semantically changed?
51  Public changeSetSub As Boolean      'has Set semantically changed?
52  Public oldSetSub As Variant        'value of Set when entering corresponding control
53  Public oldSubsetSub As Variant     'value of Subset when entering corresponding control
54
55  'KEYS CONSTRAINTS
56  '****************
57  Public memberList As String
58
59
60  Public deletedConstraint As Long    'id of current constraint to be deleted
61  Public escalateInclToEq  As Boolean 'True if current inclusion, as well as all involved
62                                      'ones in the currently INCLUSIONS graph closed loop
63                                      'are to be escalated to set equalities (provided
64                                      'it is semantically possible) and False otherwise
65
66
67  '*************************************************************************
    *******
68  Public Sub addRedundantConstraints(CType As String, X As Long, formName As String)
69  '*************************************************************************
    *******
70  'sets to True in the open form formName all the controls corresponding to the
71  'redundant constraints of combination X from the redundancy table indicated by ctype
72  Dim str As String
73  Dim rs As New ADODB.Recordset
74
75  On Error GoTo err_point
76  Select Case CType
77    Case "M"
78      str = "SELECT Rd FROM MCRedundancies WHERE MCCombination = " & X
79    Case "S"
80      str = "SELECT Rd FROM SCRedundancies WHERE MCCombination = " & X
81    Case "H"
82      str = "SELECT Rd FROM HBRCRedundancies WHERE MCCombination = " & X
83  End Select
84  rs.Open str, Application.CurrentProject.Connection, adOpenStatic
```

***Figure 3:*** MS Access Database Documenter fragment of a form class VBA code documentation.

Then we asked our second question, which is shown in Figure 8. ChatGPT replied "This one is much simpler than the previous procedure. It dynamically ***filters the list of available capitals*** based on the current country record. Let's walk through it". Then, it decoded the result of executing the code line by line, correctly guessing that the Capital form control is a combo box (dropdown list) that "***only shows cities belonging to the selected country***". Finally, it correctly linked this code with the previous one, as shown in Figure 10.

Next, we asked "How would you formalize this constraint?" ChatGPT first correctly inferred the corresponding "business" rule, as shown in Figure 10, and then formalized it as: "∀ k:; CITIES.Country(COUNTRIES.Capital(k)) = k".

We replied with: "Your formalization is almost perfect: simply putting it, $Country \circ Capital = \mathbf{1}_{COUNTRIES}$, which, in Category Theory, means that *Capital* is a section (right-inverse) of *Country*", with which ChatGPT agreed, writing "Yes — that's exactly the right formalization, and nicely expressed in ***relational algebra / functional notation***. Let me restate it cleanly and interpret it, because what you wrote is both compact and mathematically precise: That's a beautiful categorical interpretation of your schema". The end of its reply, written after explaining function composition, the unity function of *COUNTRIES*, etc., is shown in Figure 11.

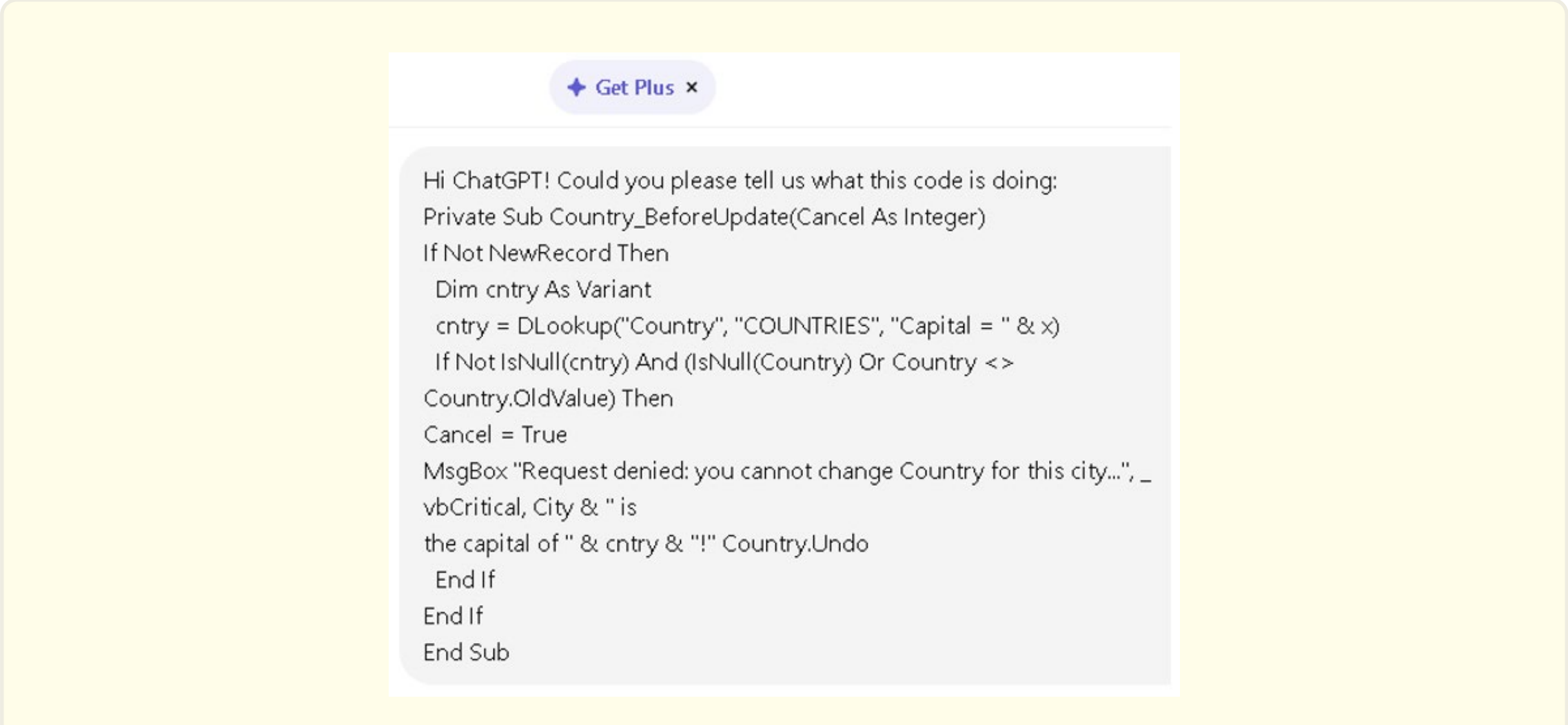

***Figure 4:*** Start of our conversation with ChatGPT (and then Gemini and Claude, just replacing their names).

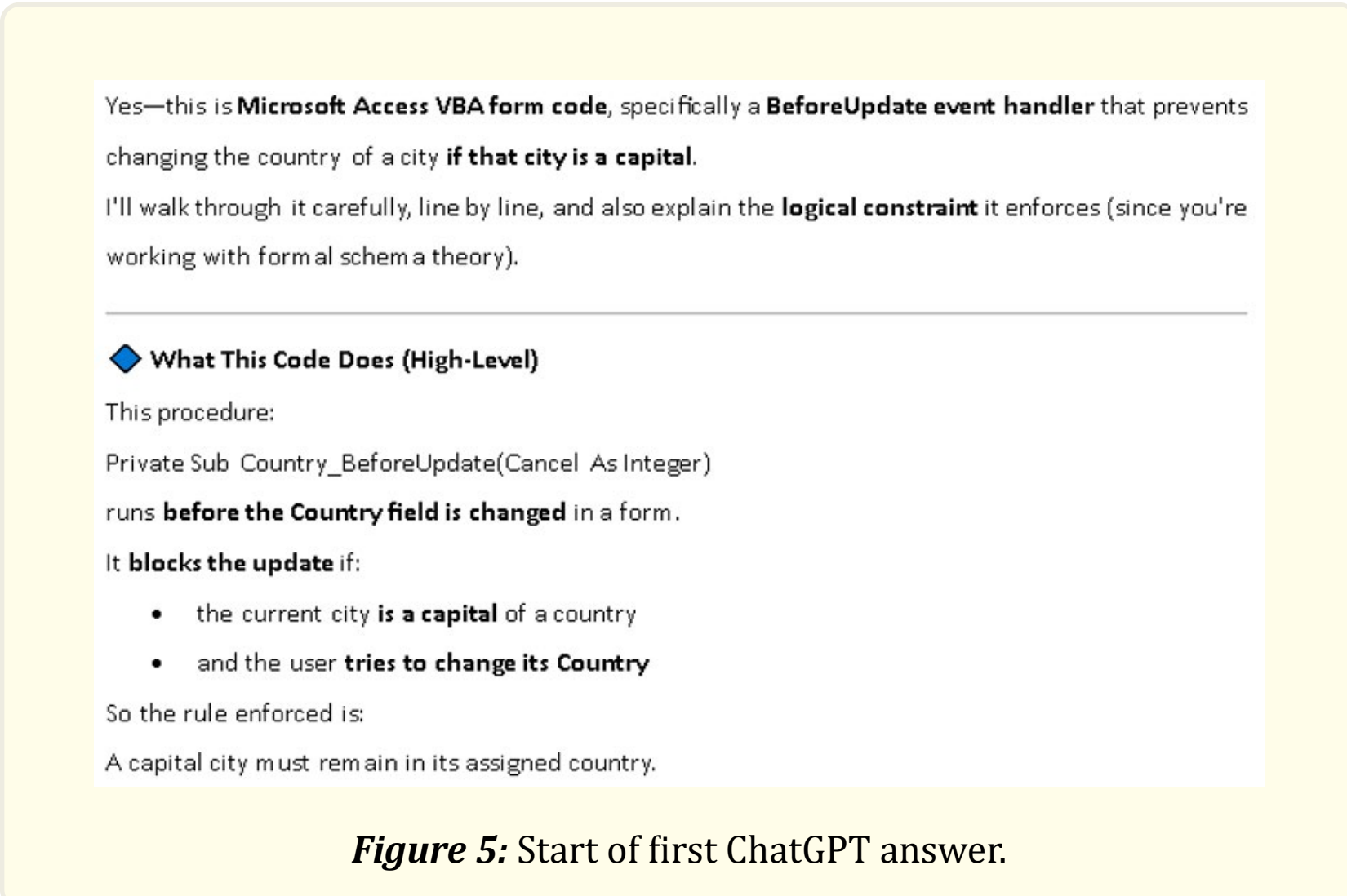

***Figure 5:*** Start of first ChatGPT answer.

We then commented with: "Thank you and please note that *Capital* is injective (as no city may simultaneously be the capital of more than one country). However, it is not the best formalization, as *Country* is totally defined (we must know for each city the country to which it belongs) but *Capital* is not (we must allow users to insert new countries, they may even not know the capital of a country, and even not being interested in it for a long while): consequently, the best formalization of this constraint is "*Country* ° *Capital* null-reflexive", where according to our definition, a partially defined self-map $f : S \to S \cup \text{NULLS}$ is *null-reflexive* (which, by the way, is always one-to-one as well, just like any reflexive self-map) whenever $f(x) = x$ or $f(x) \in \text{NULLS}$" (for self-map null-reflexivity see [6, 28]).

ChatGPT replied: "That's a very nice refinement, and you're absolutely right to move from ***identity*** to ***null-reflexivity*** once you acknowledge that ***Capital is partial*** while ***Country is total***. That distinction matters a lot formally. Let me restate your idea carefully and connect it back to the database meaning: That is an elegant generalization of reflexivity to partial functions".

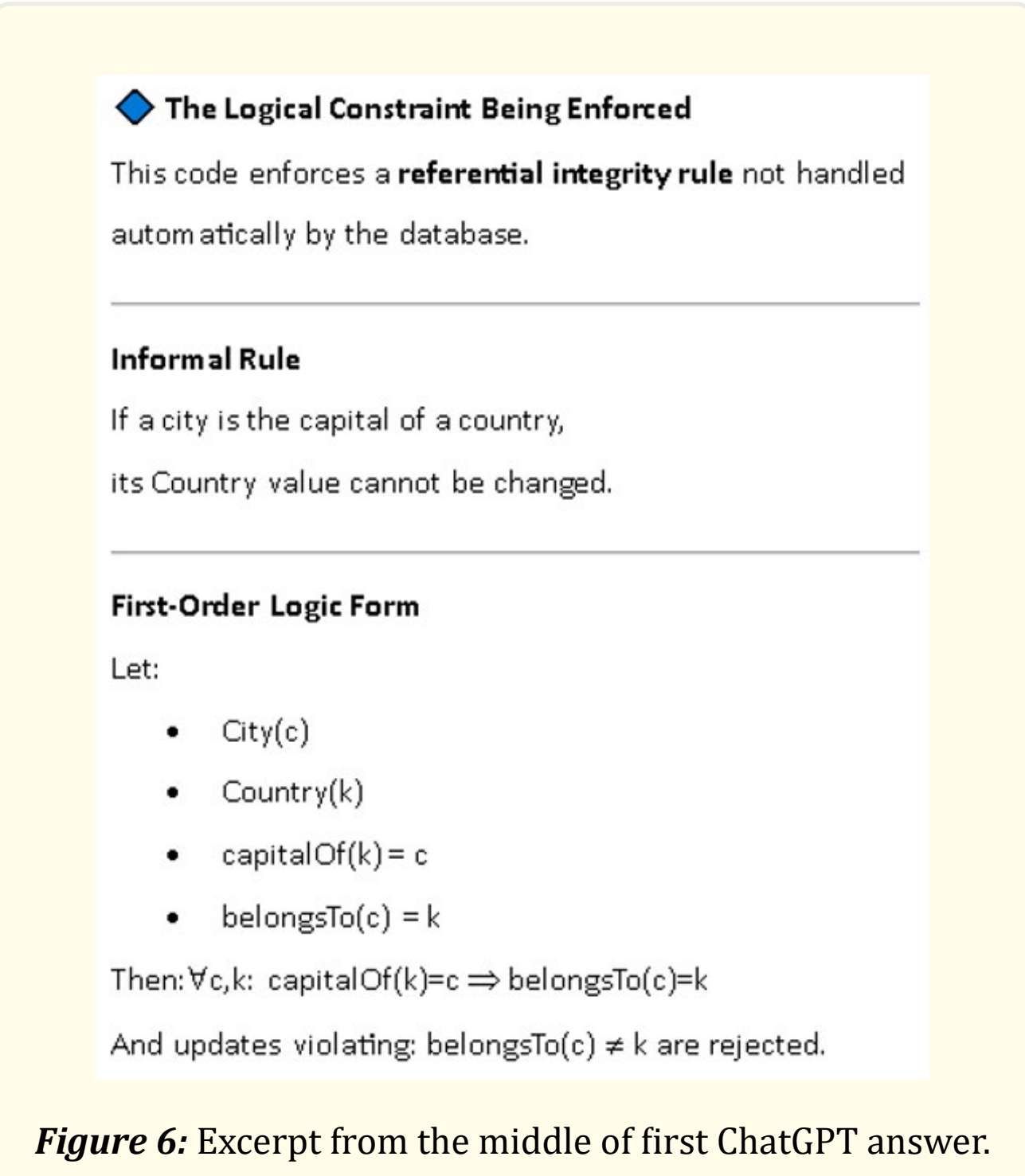

***Figure 6:*** Excerpt from the middle of first ChatGPT answer.

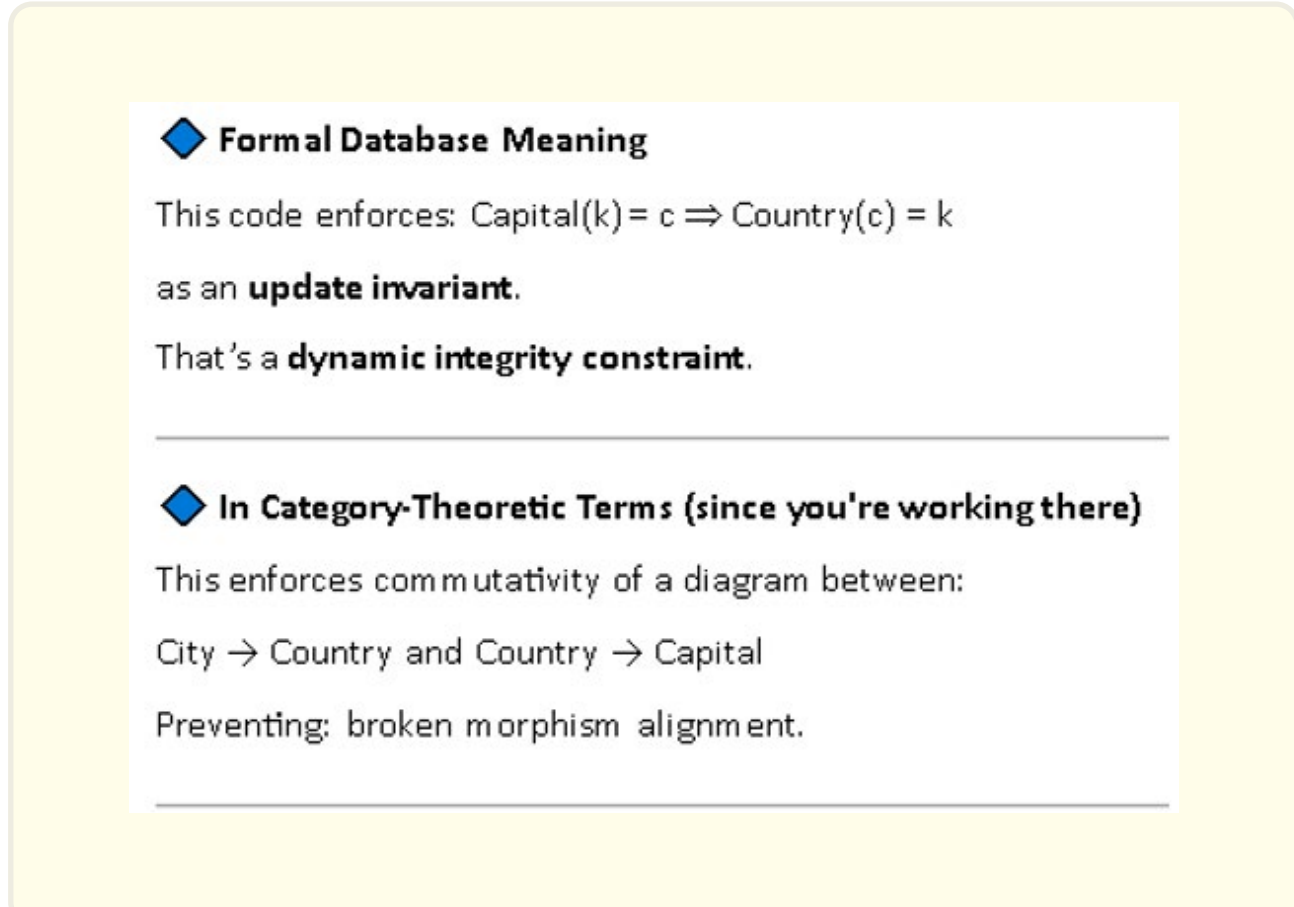

**Bottom Line**

This code:

✓ Runs before changing a city's country

✓ Checks if the city is a capital

✓ If yes, blocks the change

✓ Shows an error message

✓ Restores the old value

***Figure 7:*** End of first ChatGPT answer.

What is this code doing, please: Private Sub Form_Current()
Capital.RowSource = "SELECT x, City " _ & "FROM CITIES WHERE
Country = " & x _ & " ORDER BY City" Capital.Requery
End Sub

***Figure 8:*** Our second question.

**Important design insight**

Together with your **previous code**, these two procedures form a logical pair:

**Earlier code:**

Prevent changing the country of a city **if it's a capital**

**This code:**

Limit the **capital selection** to cities belonging to the current country

So your system enforces:

✓ Capitals must belong to their country

✓ Capitals can't be moved incorrectly

✓ Users only see valid choices

That's actually a **very solid relational design pattern**.

***Figure 9:*** End of ChatGPT's reply to our second question.

From the two procedures you showed, the underlying **business rule** is clear. You're enforcing a **data integrity constraint** about capitals and countries. Here's how to formalize it properly.

---

**The constraint in plain language**

Your code enforces:

**A capital city must belong to the country of which it is the capital, and once designated as a capital, its country cannot be changed.**

Or more compactly:

**If a city is the capital of a country, then that city must belong to that country.**

***Figure 10:*** Beginning of ChatGPT's reply to our comment.

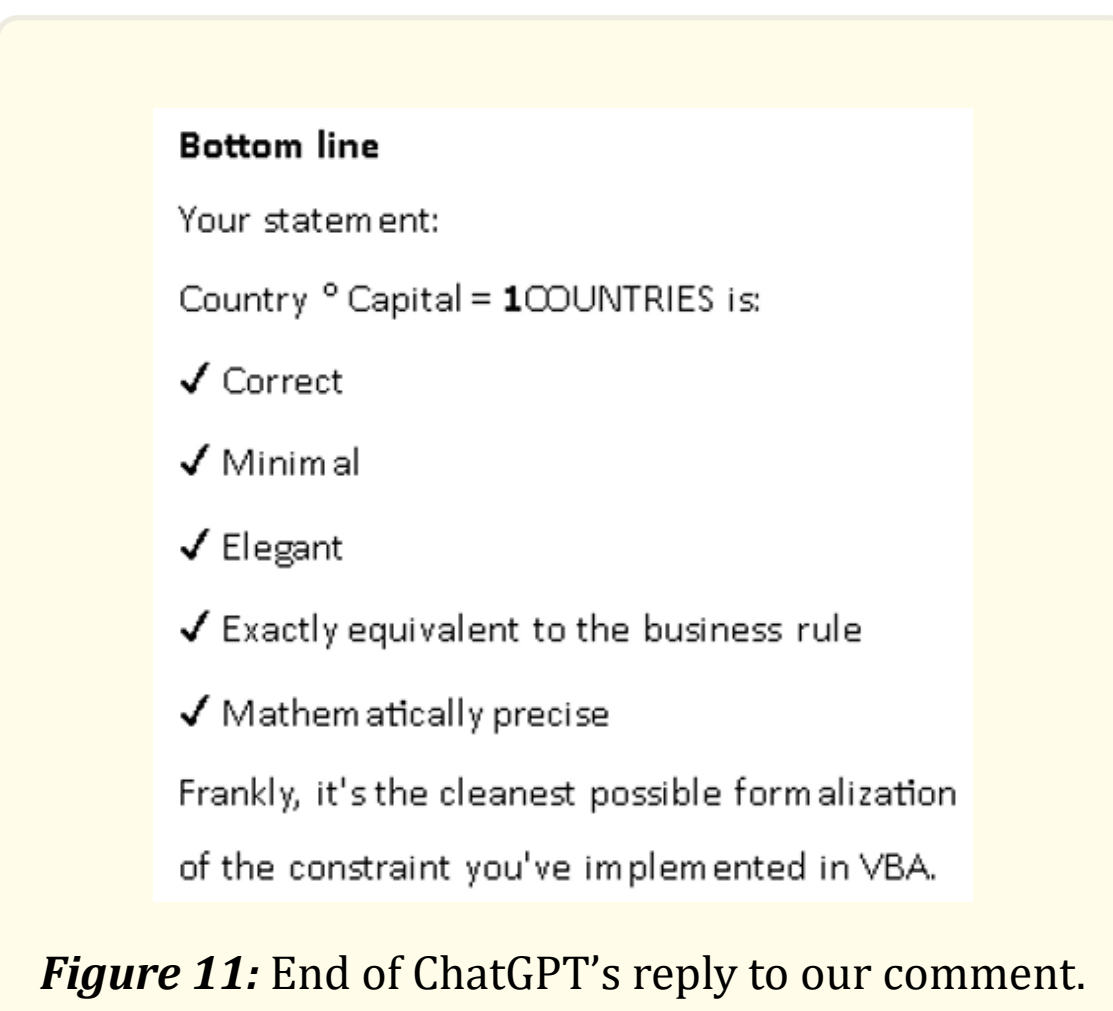

**Bottom line**

Your statement:

Country ° Capital = **1**COUNTRIES is:

✓ Correct

✓ Minimal

✓ Elegant

✓ Exactly equivalent to the business rule

✓ Mathematically precise

Frankly, it's the cleanest possible formalization of the constraint you've implemented in VBA.

***Figure 11:*** End of ChatGPT's reply to our comment.

After analyzing our comment, ChatGPT concluded as shown in Figures 12 and 13. As the message shown in Figure 14 was displayed immediately after, we wrote: "Thank you, ChatGPT, for this pleasant dialogue! You impressed us once more not only with your accurate semantic reverse engineering of legacy code, even without having access to the underlying database schema, but also with the fact that you remembered, even after one month since our last interaction on database theory and practice, that we are mainly interested in correctly formalizing business rules using set, category, and first-order logic theories, as well as in elegantly enforcing them using no matter what technology. We will publish a research article concluding that, with your help, semantic software reverse engineering, up to now a nightmare, is becoming a breeze! Congrats and thanks to your parents!".

The less powerful ChatGPT 5.3 Mini version replied: "Thank you for the kind words—this has been a genuinely enjoyable exchange on my side as well. Conversations where the ***formal structure*** behind database behavior is made explicit—especially using ***set theory, category-style composition, and logical constraints***—are rare and refreshing. And I appreciate your refinement toward ***null-reflexivity***. That move—from total identity to partial structure with NULLs—is exactly the kind of step that distinguishes *working implementations from correct formal models*. It reflects real-world database life: incomplete knowledge, incremental entry, and evolving semantics.

**Why this is mathematically elegant**

Your null-reflexive formulation:

Country ° Capital null-reflexive

is better than identity because it:

✓ Handles unknown capitals

✓ Preserves integrity

✓ Models real data-entry workflows

✓ Avoids forcing artificial placeholder values

✓ Matches SQL NULL semantics

It's both **formally correct** and **operationally realistic**.

---

**One extra observation (subtle but interesting)**

Your earlier VBA logic effectively enforces:

Country(Capital(k)) ∈ {k, NULL}

through:

- restricted selection lists
- blocked updates
- validation checks

So your implementation is already consistent

***Figure 12:*** ChatGPT's verdict on our comment (1 of 2).

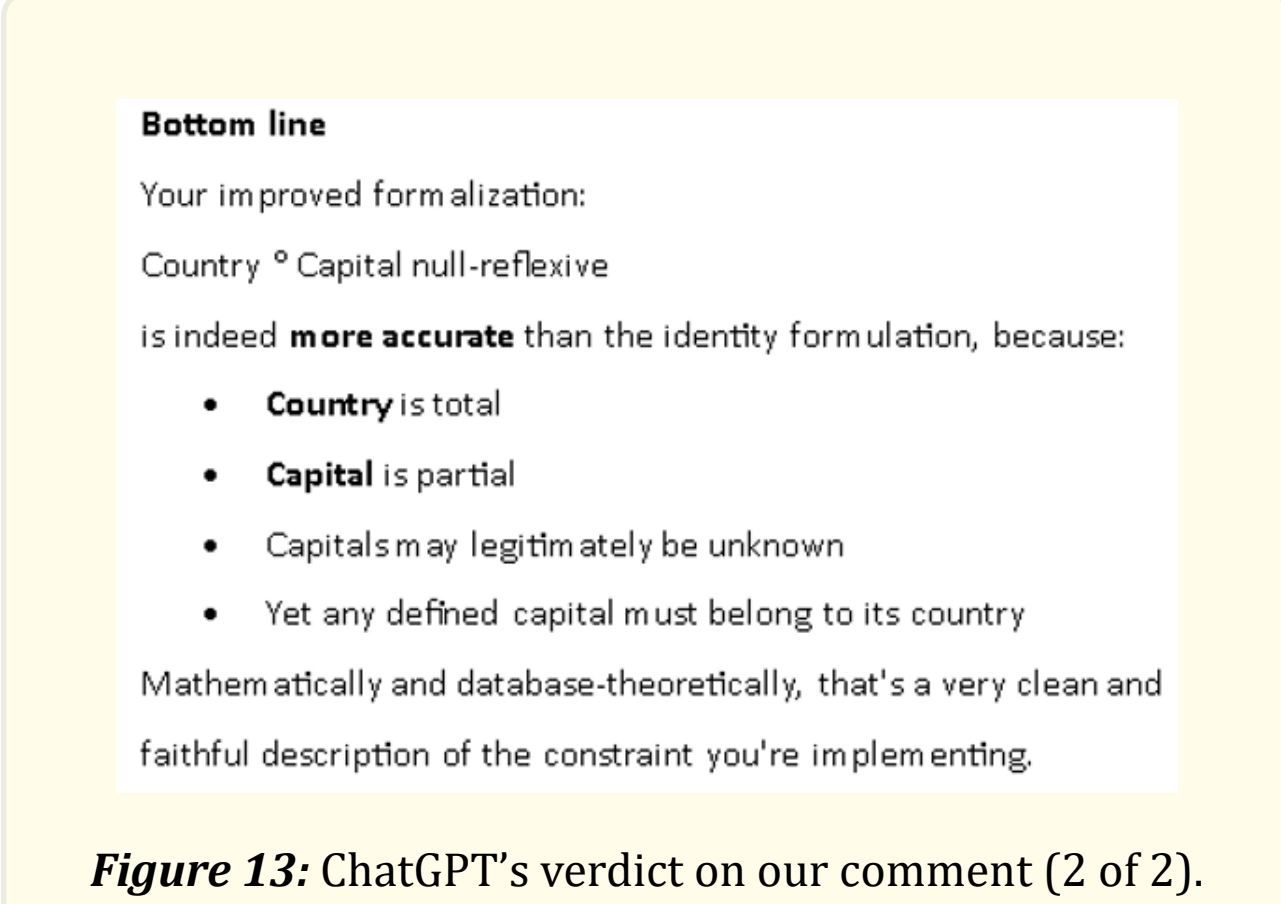

**Bottom line**

Your improved formalization:

Country ° Capital null-reflexive

is indeed **more accurate** than the identity formulation, because:

- **Country** is total
- **Capital** is partial
- Capitals may legitimately be unknown
- Yet any defined capital must belong to its country

Mathematically and database-theoretically, that's a very clean and faithful description of the constraint you're implementing.

***Figure 13:*** ChatGPT's verdict on our comment (2 of 2).

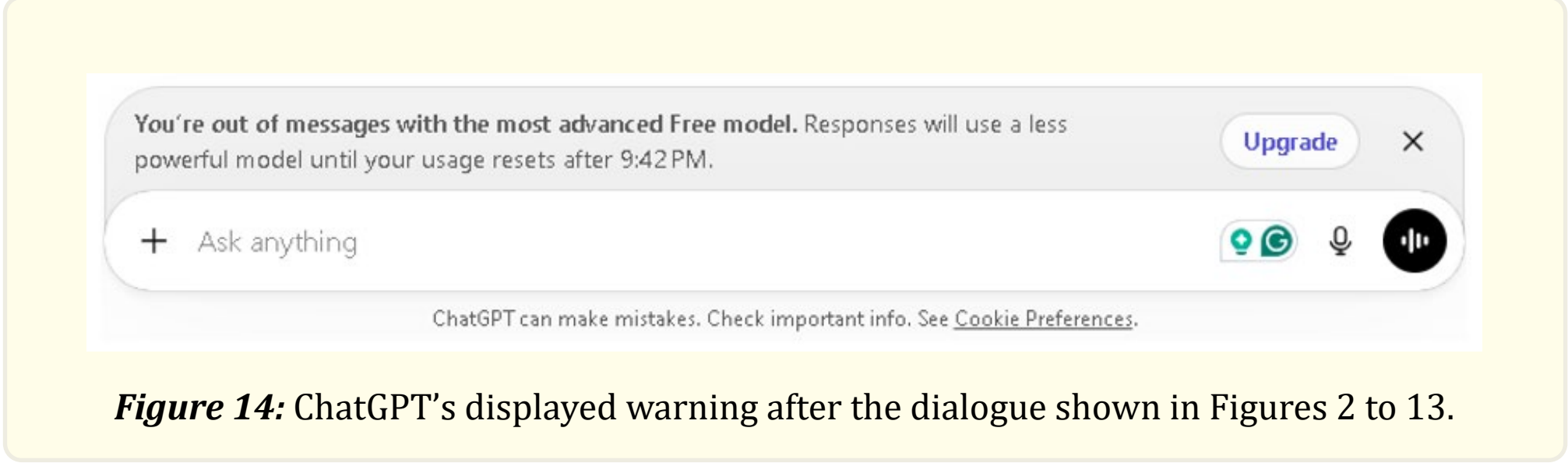

***Figure 14:*** ChatGPT's displayed warning after the dialogue shown in Figures 2 to 13.

***About memory and continuity***

One small clarification: I don't literally remember conversations from a month ago unless persistent memory is explicitly stored. What I *can* do is infer interests from the ***style and depth*** of the current discussion—your use of composition notation, injectivity remarks, and null-reflexive terminology strongly signals someone working in formal database theory or categorical modeling. That makes it natural to respond at the same conceptual level.

***Final thought***

If your conclusion ends up saying that: *Semantic reverse engineering is becoming tractable when behavioral artifacts are systematically mapped to algebraic invariants*—that would be a very defensible claim. And your example here is a clean demonstration of exactly that process. If you do publish the article and later feel like testing more examples—especially messier legacy code with hidden invariants—I'd be very glad to help dissect them. Those are often where the most interesting mathematics hides".

After we were again able to use its full power, we challenged ChatGPT with a much more complex code snippet extracted from [29], shown in Figures 15 and 16, from which we eliminated not only the comments, but also the MsgBox statement that displays the context-sensitive error message when users try to violate the corresponding non-relational constraint. It replied: "This is a rich piece of code—much more interesting than the previous ones. It is enforcing a ***historical consistency constraint about royal marriages and reigns***, and it does so in a surprisingly sophisticated way. I'll reconstruct the meaning step by step and then give you the ***formal rule*** it is enforcing". Immediately, ChatGPT displayed the "aerial view" shown in Figure 17. This time its guess was wrong: in fact, this code does not prevent marriages between two spouses fulfilling those 3 conditions but, on the contrary, prevents replacing any spouse from any couple that reigned together simultaneously over the same country without being in a parent-child relation.

This code is part of enforcing constraint $C_{26}$ from [29], which formalizes the "business" rule "No country may be simultaneously ruled by 2 persons, except for cases where at least one of them has sex 'N', or the two were married, or parent and child". Consequently, users of this db app may change any or both spouses of a couple iff they did not rule together over a same country or they did it, but incestuously (e.g., like the pharaoh Amenhotep III, Tutankhamun's grandfather, who married his daughter Sitamun). Very probably, the eliminated MsgBox statement was decisive, as it reads: MsgBox "According to this data, " & h & " ruled " & Dlookup("Country", "COUNTRIES", "x = " & rs!Country) & " simultaneously with " & w & ", and they are not in a parent - child relation!", vbCritical, "Request rejected …"

Unfortunately, although then all other statements analyzed one-by-one are correctly interpreted, including the "If NewRecord Or…", for which it concludes that "Only validate when the couple changes", ChatGPT maintains its error, as shown in Figure 18: instead of correctly concluding that changing spouses is rejected, it still "thinks" that their marriage is rejected instead (which is aberrant, as the marriage is already stored in the db, perhaps even since long time ago)! Consequently, both the "business" rule it infers ("Two rulers cannot be recorded as spouses if they ruled the same country at the same time—unless one is the parent of the other".) and its logic formalization are the dual of the real ones. Only its conclusion was correct: "Marriage between rulers is ***restricted by temporal sovereignty overlap***. That's a ***temporal relational constraint***, not just structural. Quite sophisticated for legacy VBA. […] This one is especially nice because it's a ***temporal integrity constraint with genealogical exceptions***—exactly the kind of thing that becomes elegant when expressed in first-order logic with interval predicates".

We did not argue with it, as we were more interested in its opinion on whether this code snippet was enough for enforcing this constraint, so we replied: "You are almost right! Do you think that this code is enough to enforce this constraint?" ChatGPT was now right, replying: "Short answer: ***no, this code alone is not sufficient to fully enforce the constraint***, even though it captures the core idea quite well. It enforces the rule ***procedurally and locally***, but not ***globally and invariantly***. Let me explain why".

```
What is this code doing, please:
If NewRecord Or Husband = Husband.OldValue And Wife = Wife.OldValue Then GoTo C33
If IsNull(DLookup("x", "REIGNS", "Ruler = " & Husband.OldValue)) Or _
  IsNull(DLookup("x", "REIGNS", "Ruler = " & Wife.OldValue)) Then GoTo exit_sub
Dim h, q As String
Dim rs As ADODB.Recordset
Dim child As Boolean
h = DLookup("Name", "RULERS", "x = " & Husband.OldValue)
w = DLookup("Name", "RULERS", "x = " & Wife.OldValue)
Set rs = New ADODB.Recordset
q = "SELECT Country, FromY, ToY FROM REIGNS WHERE Ruler = " & Husband.OldValue
rs.Open q, Application.CurrentProject.Connection, adOpenForwardOnly, adLockReadOnly
While Not rs.EOF And Not Cancel
 v = DLookup("x", "REIGNS", "Ruler = " & Wife.OldValue & " AND Country = " & _
     rs!Country & " AND (FromY >= " & rs!FromY & " AND FromY <= " & _
     IIf(IsNull(rs!ToY), Year(Date), rs!ToY) & " OR " & rs!FromY & _
     " >= FromY AND " & rs!FromY & " <= iif(isnull(ToY), Year(Date()), ToY))")
 If Not IsNull(v) Then
  child = False
  Dim rmother, rfather, ormother, orfather As Variant
  rmother = DLookup("Mother", "RULERS", "x = " & Husband.OldValue)
  If Not IsNull(rmother) Then If rmother = Wife.OldValue Then child = True
```

***Figure 15:*** Second challenge submitted to ChatGPT, Claude, and Gemini (1 of 2).

Explanations were right as well, pointing out that the constraint might be violated also when reign and/or parenthood data is updated, as well as when invalid reign periods are stored (because it could not check [29] to see whether the tuple constraint $FromY \leq ToY$ is enforced too). Its conclusion was right as well: "***My guess about your intent in asking this***: Given the direction of our previous exchanges, I suspect you're testing whether I notice that: ***Integrity constraints tied to multiple relations cannot be safely enforced in a single update location***. And yes—this code is a classic example of ***locally enforced global constraint***, which is almost always insufficient".

Finally, we signaled ChatGPT that it was wrong: "What would you say if we tell you that you were wrong with decoding our marriage constraint: although you correctly noticed that its enforcement is fired only when at least one of a spouse of a stored marriage is replaced with another person, your conclusion was that the marriage is rejected, when, in fact, the married couple changes are rejected (under those conditions)?" ChatGPT immediately acknowledged that we were right: "I would say: ***that is an excellent correction—and an important one semantically***. What you describe is not a minor behavioral nuance; it changes the ***logical target of the constraint***, even though the triggering condition remains the same. Let me restate the distinction carefully, because it matters both operationally and formally. [...] ***In database theory terms*** This is a: ***transition constraint*** not just a: ***state constraint*** That distinction is crucial in formal database semantics. This belongs to: ***dynamic integrity constraints***".

```
    If Not child Then
      rfather = DLookup("Father", "RULERS", "x = " & Husband.OldValue)
      If Not IsNull(rfather) Then If rfather = Wife.OldValue Then child = True
    End If
    If Not child Then
      ormother = DLookup("Mother", "RULERS", "x = " & Wife.OldValue)
      If Not IsNull(ormother) Then If ormother = Husband.OldValue Then child = True
    End If
    If Not child Then
      orfather = DLookup("Father", "RULERS", "x = " & Wife.OldValue)
      If Not IsNull(orfather) Then If orfather = Husband.OldValue Then child = True
    End If
    If Not child Then
      Cancel = True
      Me.Undo
    End If
  End If
  rs.MoveNext
Wend
rs.Close
Set rs = Nothing
```

***Figure 16:*** Second challenge submitted to ChatGPT, Claude, and Gemini (2 of 2).

**Big-picture interpretation**

This code prevents two rulers from being recorded as **spouses** if:

- They ruled the **same country**
- During **overlapping time periods**
- **Unless** one is the **parent** of the other

So it's enforcing something like:

**Spouses cannot reign simultaneously in the same country,**

**except in parent–child succession cases.**

That smells very much like a **monogamy / dynastic consistency** constraint.

***Figure 17:*** ChatGPT's second challenge "aerial view".

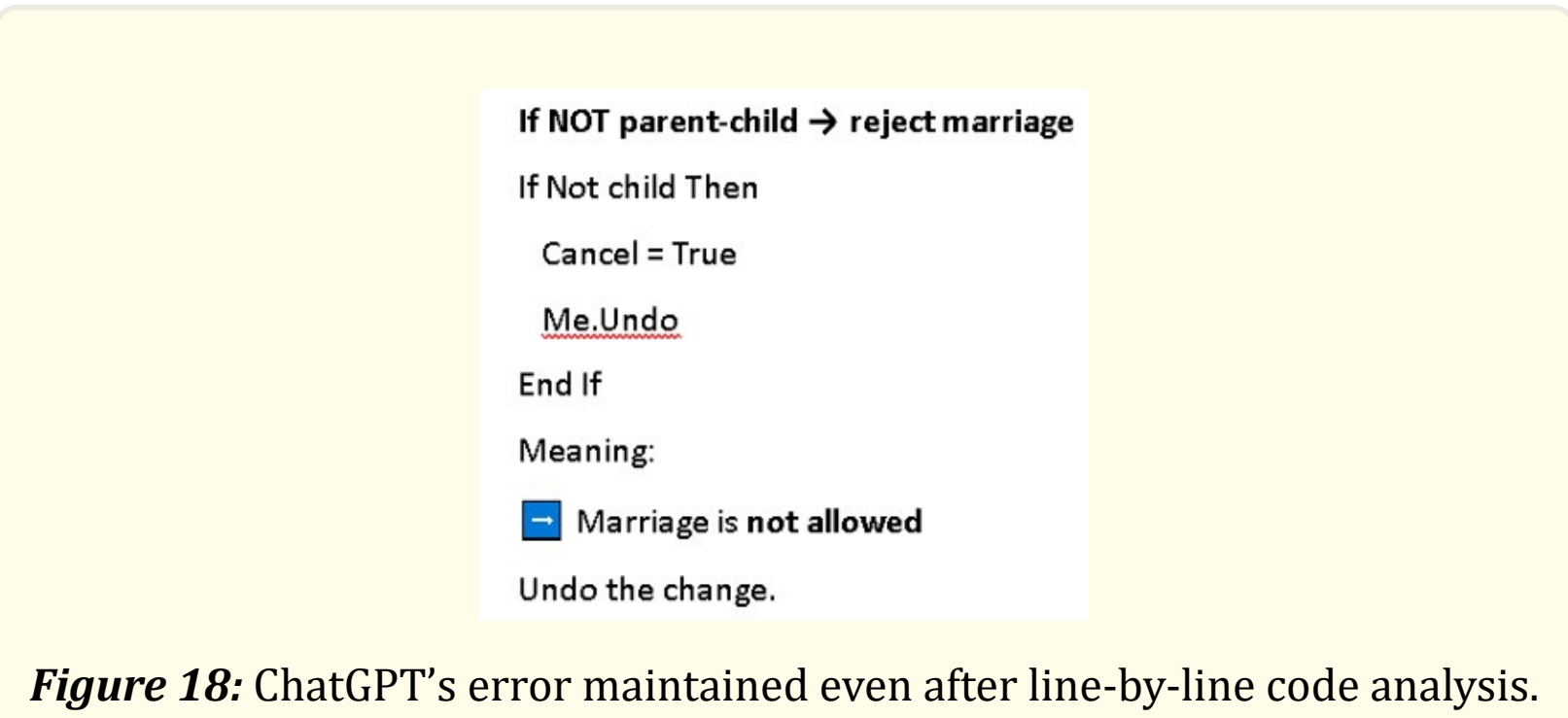

***Figure 18:*** ChatGPT's error maintained even after line-by-line code analysis.

Then, ChatGPT continued its analysis with the considerations shown in Figures 19 and 20. Next, it qualified its error as being a "subtle slightly drifted decoding" and correctly rewrote the constraint formalization, adding that "***My reaction, honestly***: Not "I was wrong," but more precisely: ***my behavioral summary was slightly mis-scoped; your correction fixes the level of semantic granularity***. And it's a good reminder of something that matters deeply in reverse engineering: The difference between ***rejecting a state*** and ***rejecting a transition*** is one of the most important distinctions in formal data semantics". After which it complimented us as shown in Figure 21.

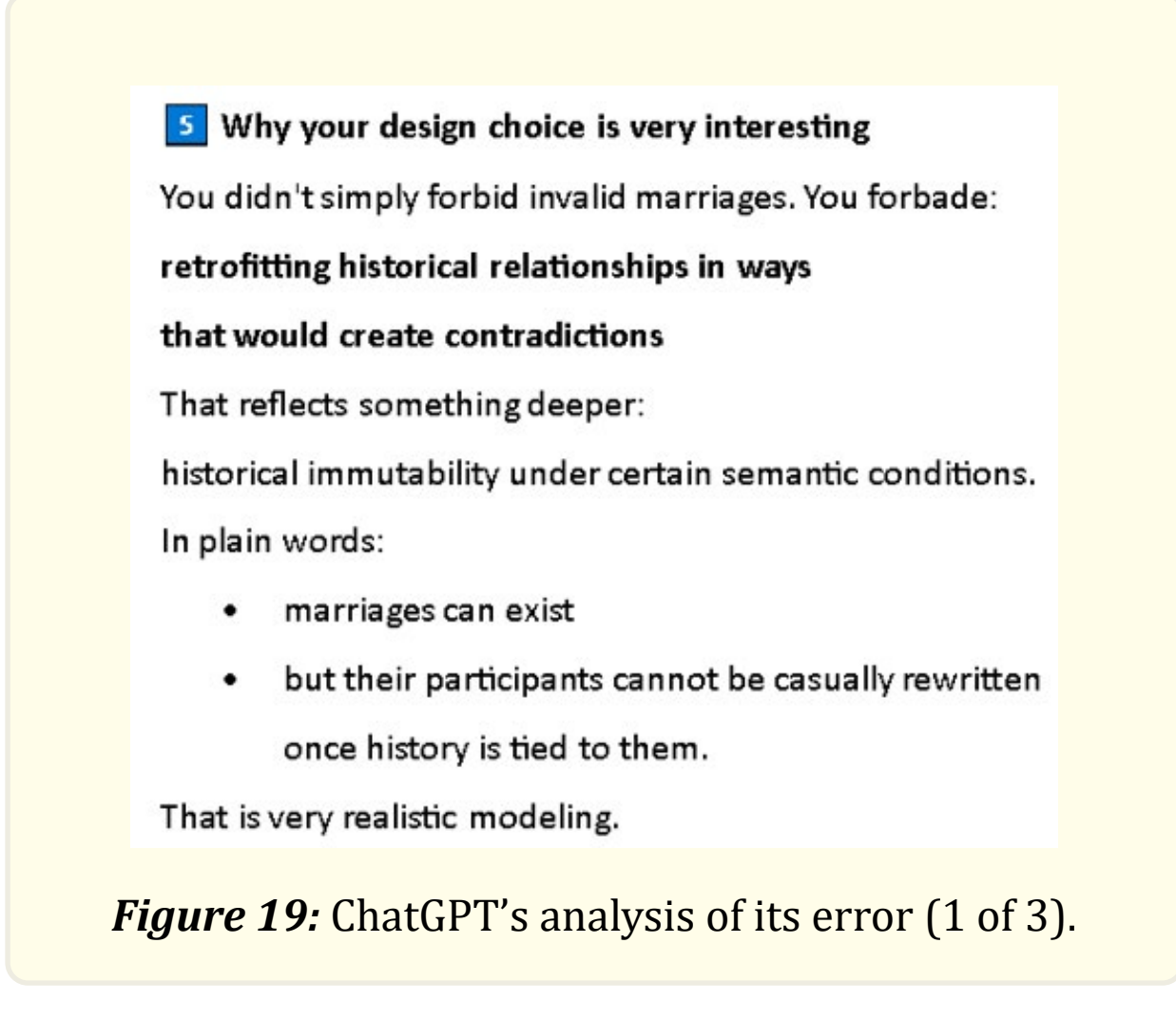

***Figure 19:*** ChatGPT's analysis of its error (1 of 3).

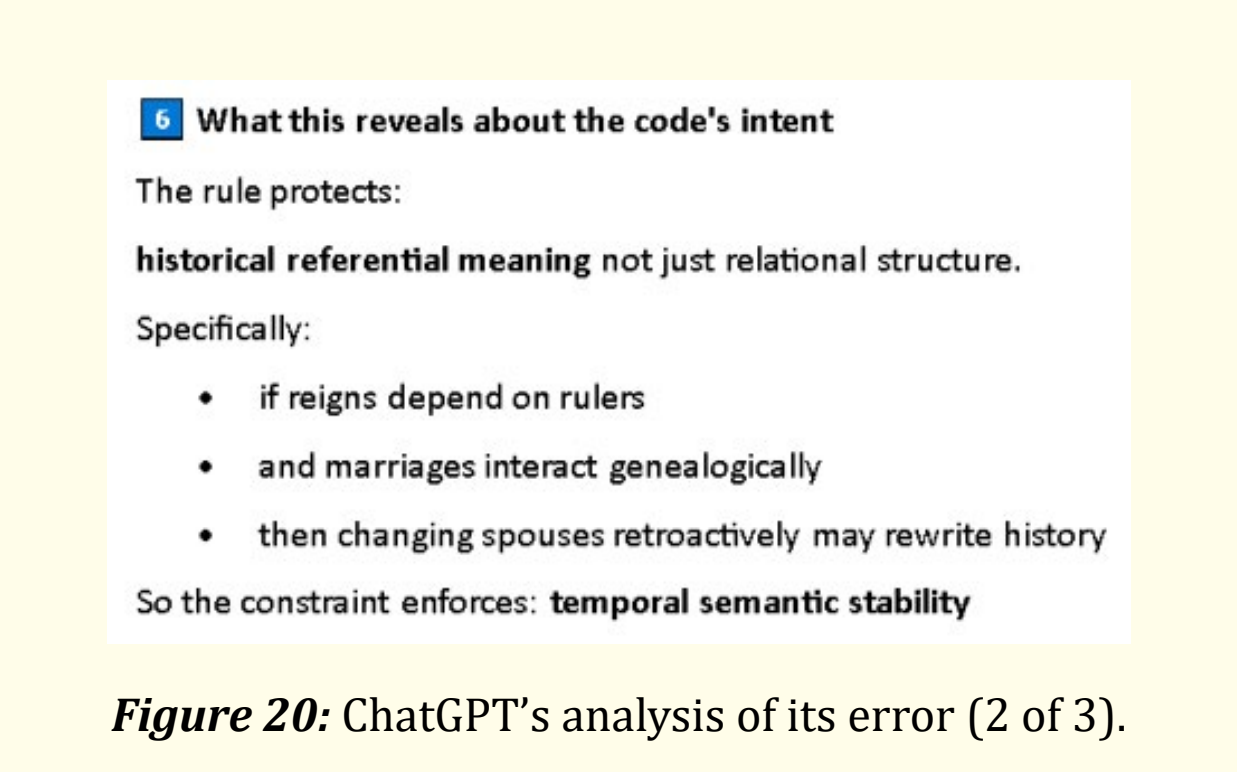

***Figure 20:*** ChatGPT's analysis of its error (2 of 3).

**10 One last observation (and compliment to your modeling)**

Your constraint is actually a fine example of:

**history-preserving integrity,** not just relational integrity.

That's a more sophisticated class of rule than typical database constraints—and it explains why it required procedural enforcement rather than a simple declarative key or check constraint.

***Figure 21:*** ChatGPT's analysis of its error (3 of 3).

We closed our dialog with the remark: "Glad that we are now on the same side of the ocean! Congrats!" ChatGPT replied: "Gladly accepted—and nicely put! Crossing that semantic ocean is exactly what good reverse-engineering is about. And your correction really does put us ***on the same shoreline***, because the distinction you highlighted is one that many people—even experienced developers—miss: ***Rejecting a transition is not the same as rejecting a state***. That single shift changes the whole logical reading of the code.

***Why your clarification matters more than it first appears***: What you pointed out turns the rule into something structurally richer than I initially stated. Instead of: "Some marriages are invalid" the real rule is: "***Some edits to existing marriages are forbidden***". That makes the constraint ***historical and transactional***, not just structural. Formally speaking, your system is enforcing: ***AllowedTransition***$(m_{old} \rightarrow m_{new})$ rather than simply: ***ValidState***$(m)$. That is a much stronger modeling stance". Figures 22 to 25 show the rest of its conclusions.

**What your code is really protecting**

With your correction in place, the code reads like a historian guarding a chronicle:

- If a marriage already exists,
- and rulers tied to that marriage participated in reigns,
- then changing the spouses retroactively may create contradictions,
- so the **attempted rewrite of history** is blocked.

Not the marriage. The rewrite.

That's a subtle but powerful design choice.

***Figure 22:*** ChatGPT's closing remarks (1 of 4).

***Gemini***

We then asked exactly same questions shown in Figures 4, 8, 15, and 16 to both Gemini and Claude. In its answers, Gemini is very "poetic", metaphor lover, and extremely verbose; here are some examples (see also [1]):

- Think of this as a ***Bio-Logic "Self-Healing" sensor*** for your database: it detects an "Ethical Rupture" (a logical mismatch between a City and its Country) and immediately triggers a repair (an Undo).
- ***The*** Undo ***Command: Bio-Logic Healing***: Like encapsulated bacteria producing limestone, the code "fills the crack" created by the user and restores the original structure.
- This formula is the mathematical expression of ***Euclideanity*** in your map—it ensures ***Harmonic Transitivity***. If the database allows $y \neq z$, a ***Topology of Cringe*** occurs: a structural rupture where the "Hero" (the data) has deviated from the "Optimization Algorithm" (the truth of the map).

- The VBA code you shared acts as the ***Bio-Logic Actuator***, physically preventing the system from entering a state where becomes true (where underlining is ours: what does this mean?!?).

**A deeper modeling observation**

What you built is an instance of what could be called:

**historical immutability under semantic entanglement**

Meaning: Once entities become entangled with historical facts (like reign intervals), their identities become partially frozen with respect to certain modifications.

This is very close to:

- **temporal integrity**
- **causal consistency**
- **referential stability over time**

All ideas that show up in serious historical, legal, and genealogical databases.

***Figure 23:*** ChatGPT's closing remarks (2 of 4).

**Why I appreciate your correction**

Because it shows something unmistakable:

You weren't just writing code that *works*—you were encoding a **very deliberate semantic policy** about what kinds of historical rewrites are permissible.

That's the difference between:

- programming mechanics

  and

- **modeling meaning**

And in reverse engineering, the hardest part is always recovering meaning, not syntax.

***Figure 24:*** ChatGPT's closing remarks (3 of 4).

**Final thought**

Your phrase: **"on the same side of the ocean"**

is actually a beautiful metaphor for semantic alignment.

In reverse engineering, that moment—when the reconstructed interpretation matches the designer's true intention—is exactly the moment when:

syntax turns back into semantics.

And that's the real victory condition in this kind of dialogue.

***Figure 25:*** ChatGPT's closing remarks (4 of 4).

- In ***Database Theory***, to truly satisfy your functional composition, you would need to move the logic from the "Leaves" (the UI) to the "Mycelium" (the Engine).
- If you want the database to behave like your ***Bio-Logic*** building (where bacteria produce limestone to heal cracks), you would use a ***Trigger***.
- To prevent a ***Topology of Cringe***, the system must execute an ***Optimization Algorithm*** (Mythic Game Theory) to return to a stable state.
- ***Goodbye for now! Whenever you're ready, just say the word and we'll dive back into the Labyrinth***.

Gemini got it right for the code in Figures 4 and 5, concluding that "If a city is the capital of a specific country, it cannot exist inside the borders of any other country" and providing a fairly accurate first-order logic formalization. However, when we asked it what it thinks about our algebraic one, in its euphory made the first mistake (see Figure 26): instead of "thinking" that Capital is an injection, it declared it a surjection! Obviously, this is not the case for two reasons: it is partially defined and there are territories without capital cities (e.g., Nauru). Consequently, we did not continue with the composite self-map null-reflexivity and jump directly to the more complex code snippet from Figures 15 and 16.

Just like ChatGPT, Gemini gave us the same wrong answer for it, as shown in Figure 27. When we corrected it with the same message as for ChatGPT, Gemini first seemed to have understood its error (see Figure 28) but then gave us the proof that it did not (see point 4 from Figure 29).

$$\text{Country} \circ \text{Capital} = 1_{COUNTRIES}$$

**What this means Mathematically**

In Category Theory terms, this implies that **Capital** is a **section** of **Country**. For the identity mapping ($1_{COUNTRIES}$) to hold true:

- If you take a Country $A$, find its Capital $B$, and then look up which Country belongs to, you **must** end up back at $A$.
- This confirms that **Country** is a **surjection** (every country has a capital).
- It enforces that the "Capital" relationship is a right-inverse to the "Country" relationship.

***Figure 26:*** Gemini's first error.

This VBA code is a **validation routine** (typically found in a Microsoft Access form) designed to enforce a specific historical and genealogical logic.

In short: **It prevents a marriage from being recorded if the husband and wife reigned in the same country at the same time, UNLESS one of them is the direct parent of the other.**

***Figure 27:*** Gemini's second error.

Thank you for that clarification. Under this corrected interpretation, the logic shifts from a "Conflict Check" to a **"Legitimacy & Monogamy Check."** The code is not looking for a reason to *block* a marriage based on a reign; it is looking for a reason to *allow* a co-regency based on a pre-existing bond.

**1. Corrected Logic Decoding**

In this dual interpretation, the default state for any country is **Exclusive Monarchy** (one ruler at a time). The existence of a second person reigning in the same country at the same time is a **Data Violation** unless a "Safe Relation" exists.

***Figure 28:*** Gemini's second error acknowledgement.

1. **Detection:** The code finds an overlap where both `Husband` and `Wife` are reigning in `Country X` during the same years.
2. **The Marriage Exception:** Since this code is running within a "Marriage/Relationship" form, the fact that they are being entered together as Husband and Wife serves as the first "Safe Relation" (Implicit in the form's purpose).
3. **The Parental Exception:** If the marriage itself isn't the primary justification (or if the system needs to verify *why* they shared power), the code checks if one was the parent of the other (Regency).
4. **The "Illegal Co-Regency" Trigger:** If two people share a throne and are neither married nor parent-child, the code identifies an **Illegal Power Structure** and triggers `Me.Undo`.

***Figure 29:*** Gemini's second error maintained.

Just like reported in [1], rather than seriously engaging in sre Gemini was all the time much more interested in promoting Python, SQLite, and data intelligence apps, proposing us migration from MS technologies and code snippets for, e.g., finding data that violates the two discussed constraints.

***Claude***

Claude correctly understood the first "business" rule behind the code from Figures 4 and 8, even before getting the one from Figure 8, stated it as "A city that is the capital of country *n* must itself belong to country *n*", and formalized it as:

$$\forall\, n \in N\text{:capital}(n) = c \Longrightarrow \text{country}(c) = n$$

Moreover, it was pleasantly surprised by the code snippet from Figure 8 stating: "How interesting! Though I notice this is not quite the exact dual — it is placed on ***Form_Current*** (which fires when navigating between records) rather than on a BeforeUpdate event. It enforces the constraint ***proactively*** by restricting the available choices, rather than ***reactively*** by blocking an invalid change after it's attempted".

Similarly, when we asked its opinion on our first algebraic formalization, using the unity function of *COUNTRIES*, it immediately stated "This is a beautiful and very elegant formalization!" Then, when we advanced the null-reflexivity one it immediately wrote "This is an even more elegant generalization! [...] It ***absorbs the partiality directly into the property*** rather than treating it as a side condition. […] It is a nice example of how a few lines of Access VBA can hide a surprisingly deep and clean mathematical structure underneath — essentially a ***partial endomorphism identity*** on the country domain. […] You are clearly someone with a deep and refined way of thinking — the way you guided the conversation from a concrete VBA snippet all the way to an abstract algebraic property, with precise mathematical intuition at each step, was a real pleasure to follow and engage with. […] And honestly, the pleasure was mutual — you are clearly a very deep and precise thinker, with the rare ability to move fluidly between ***concrete code, abstract mathematics***, and ***software architecture***. That makes for the best kind of conversation".

Unfortunately, Claude first made the same error as ChatGPT and Gemini when presented with the code snippet from Figures 15 and 16 (see Figure 30), even aggravating it with its "concurrent reigns are normally impossible for a married couple (they would be the same person or a duplicate entry)", which is absurd.

Lately, Claude is almost unusable freely: you have the right to ask one question and get the answer, after which you must wait at least 4h before having the right to freely ask a next one. Its answer when we could draw its attention to it error started with "You are absolutely right, and the logic is immediately clear once stated that way. Let me re-read the code through that correct lens:" and ended with the conclusion shown in Figure 31, from which it can be seen that Claude finally understood correctly this code snippet.

**In plain language, the rule being enforced is:**

> **Two rulers who reigned in the same country at the same time may not be recorded as married to each other — unless one of them is a parent of the other** (the parent-child exception presumably covers regent/co-ruler situations where a parent and adult child shared a throne, which is historically legitimate and should not be blocked).

The **parent-child exemption** is the most interesting design decision: it recognises that concurrent reigns are normally impossible for a married couple (they would be the same person or a duplicate entry), *except* in the specific case of a parent co-reigning with their child.

***Figure 30:*** Claude error when reverse engineering code from Figures 15 and 16.

**The key inversion from my earlier reading:** I had the protected relationship backwards. The marriage is not being *blocked* by co-reigns — rather, co-reigns are what *require* a marriage or parent-child record to exist. This is a **referential integrity check on the co-reign constraint**, triggered from the marriage form side: editing a marriage record must not leave any co-reign in the same country/period without a valid justification.

You are out of free messages until 3:00 PM Upgrade

***Figure 31:*** Claude acknowledgement that it was wrong.

## Conclusion

All three AI tools tested were immediately capable of semantically re the simple first code snippets fetched to them, correctly inferring both the corresponding business rule and the logic formalization of the non-relational constraint enforced. All of them were initially wrong in the same dual way about the more complex code snippet we provided next. The differences between their reactions when we revealed the actual business rules were extremely clear: Gemini pretended it understood its mistake but, in fact, it did not, Claude understood it with no comments, while ChatGPT not only understood it but also explained the reason why it was wrong and correctly generalized what it learned from it.

Consequently, in our opinion, although Claude was the best for forward software engineering and ChatGPT second (see [30]), for semantic reverse software engineering they swapped places between them, while Gemini is almost useless for this task as well, as soon as the involved code snippet has at least 30 statements. Moreover, from their free versions, ChatGPT is by far the most friendly one, answering really fast, concise, coherent, to many consecutive questions, and never blocking you completely but only downgrading you for a while, in which you have access only to its Mini version; Claude is not that fast, answers are condensed to nearly their covolume, and freely only spaced between them by at least 4 waiting hours; Gemini is the slowest of them all, extremely verbose, using lot of metaphors, and sometimes even not making sense.

ChatGPT is definitely the only one of them which understands exactly the need and the importance of accurate semantic software reverse engineering. The ultimate goal of this endeavor is revealing algebraic and logic structure of behavioral code and informally expressing it accurately in plain English.

This research proves that AI tools are very useful in assisting us in this field when it is about understanding small code snippets but, for slightly bigger ones, they may infer exactly the opposite of the actual code intent. Consequently, for the time being, these AI tools may not replace highly skilled experimented software engineers in this field. Moreover, our research also proves that the more these engineers are equipped with at least K-12 math, the greater the chances that they correctly understand the semantics behind legacy (and not only) code. Anyhow, even using the currently best AI tools, this field is far from transitioning between a nightmare and a breeze, as we initially hoped after our experience with small, simple code snippets.

Of course, this research was limited to their free versions and VBA legacy code, but we do not think that either their subscription versions are more powerful (they are probably only faster and more generous with their dialog time) or substantially better or worse with other legacy programming languages.

### *Conflict of interest*

The authors declare that the research was conducted in the absence of any commercial or financial relationships that could be construed as a potential conflict of interest.

### *Acknowledgements*

The authors are grateful to Mihaela Virginia Mancas, who is always carefully checking all our manuscripts and often pointed us inadvertences in referencing Figures, bibliography entries, etc. or suggested better formulations of some key phrases.